\documentclass[twocolumn,showpacs,preprintnumbers,amsmath,amssymb]{revtex4}
\usepackage{epsfig}
\newcommand{\bm}[1]{ \mbox{\boldmath $#1$}  }

\begin{document}

\title{Isospin mixing and energy distributions in three-body decay}

\author{E. Garrido} 
\affiliation{Instituto de Estructura de la Materia, CSIC, Serrano 123, E-28006
Madrid, Spain}
\author{D.V. Fedorov} 
\author{H.O.U. Fynbo}
\author{A.S. Jensen}
\affiliation{Department of Physics and Astronomy, University of Aarhus,
DK-8000 Aarhus C, Denmark}
\date{\today}

\begin{abstract}
The structure of the second 2$^+$ resonance in $^{6}$Li is
investigated with special emphasis on its isospin 0 components.  The
wave functions are computed in a three-body model ($\alpha$+$n$+$p$)
using the hyperspherical adiabatic expansion method combined with
complex scaling.  In the decay into three free particles the symmetry
conserving short-range interaction dominates at short distance whereas
the symmetry breaking Coulomb interaction dominates at intermediate
and large distances resulting in substantial isospin mixing.  We
predict the mixing and the energy distributions of the fragments after
decay. Computations are consistent with available experiments. We
conjecture that nuclear three-body decays frequently produce such
large isospin mixing at large distance where the energy distributions
are determined.
\end{abstract}

\pacs{21.45.+v, 31.15.Ja, 25.70.Ef}

\maketitle
\paragraph*{Introduction.}

The spatial extension of halo states depends sensitively on their
binding energy \cite{jen04}.  The reason is that the outer particles
forming the loosely bound halo are confined by an attraction or a
barrier of moderate size.  This sensitivity to the energy must be
particularly important for resonances owing their existence to a
confining barrier.  For the same reason, due to the 
effect of the Coulomb interaction, the structure of isobaric analog 
resonances can be very different. This influence can also be 
reflected in mixing of different isospin components. For ordinary states the 
isospin mixing has been established to be in the range $10^{-4}$-$10^{-5}$
\cite{boh69}.  For isobaric analogs to halo states the mixing could
be substantially larger \cite{han93}.

Early theoretical investigations of decay of three-body analog halo
states in light nuclei indicated that isospin is very well conserved
\cite{suz91,ara95}. A very small isospin mixing is also deduced from
the reduced measured branching ratio of decay of the highest known 2$^+$ 
resonance in $^6$Li (with excitation energy of 5.37 MeV and associated to 
isospin 1) into the $\alpha$-deuteron two-body system \cite{deb71,wit75}
(admixture of $8\cdot 10^{-3}$ or less).
Nevertheless, isospin mixing in $^{6}$Li-states is also investigated in a number of
reaction experiments $^2$H$(\alpha,\alpha)^2$H$^*$ where $^2$H$^*$ is
the neutron-proton system in a spin singlet relative state which cannot be
populated when isospin is conserved.  Strong evidence is claimed for the formation of 
$^2$H$^*$ which means substantial isospin mixing and especially when the incident 
energy is low \cite{wer81,gai88,nie92}.  Reasonable agreement between 
measurements and simple model calculations can only be obtained with inclusion 
of both spin zero neutron-proton $s$-wave ($^2$H$^*$) and nucleon-$\alpha$ relative
$d$-waves. Some of these authors \cite{wer81}  suggest the large isospin mixture to 
be due to direct reactions bypassing the $^6$Li resonance, whereas others \cite{gai88} 
call for better methods for including the Coulomb interaction in three-body calculations.

Recent investigations emphasized that the structure of three-body
resonances can change substantially from small to large distances
\cite{gar05b,gar05c,gar05f}. This means for instance that the amount 
of isospin may vary substantially with the (hyper)radial coordinate describing 
the wave function.  In other words, the relative weights of different partial 
wave components can be very different from small to large distances, e.g. in $^6$Li
($\alpha$ + n + p) for neutron and proton in relative $s$-waves and in
triplet or singlet spin states corresponding to $T=0$ or $T=1$,
respectively.  The absence of decay of the $2^+$ resonance in $^6$Li
into the deuteron channel is not a direct evidence for a correspondingly small
admixture of isospin 0.  The dominant isospin 1 wave function at
small distance can decay into continuum final states with isospin 0
provided the coupling is sufficiently strong.  This is most likely at
distances just outside the ranges of the short-range interactions
where the Coulomb interaction still is substantial and completely
dominating.

The purpose of the present letter is to investigate the isospin
conservation in the 2$^+$ three-body resonance in $^6$Li, that is the
isobaric analog state of the corresponding 2$^+$ resonance in $^6$He 
and $^6$Be, for which isospin zero components are forbidden.
We start with a sketch of the necessary theoretical framework. We then
discuss the structure of the resonance and present the
final state energy distributions. We finish with a brief summary and the 
conclusions.

\paragraph*{Theoretical framework.}

Bound states and resonances are obtained with the hyperspherical adiabatic expansion method
combined with complex scaling. These three-body wave functions $\Psi$ are then expressed
as a linear combination of the complete set of functions
$\{\Phi_n(\rho,\Omega)\}$ \cite{nie01}
\begin{equation}
\Psi(\bm{x},\bm{y})=\frac{1}{\rho^{5/2}}\sum_n f_n(\rho) \Phi_n(\rho,\Omega) \; ,
\label{eq1}
\end{equation}
where $\rho$ is the hyperradius and the five hyperangles 
$\Omega = \{\alpha,\Omega_x,\Omega_y\}$ can formally be chosen in any of
the three Jacobi sets, or as we prefer the wave function can be
given in terms of three Faddeev components each associated with one
Jacobi set.  Each of these Faddeev components are in turn expanded in
partial waves with a basis of corresponding hyperspherical harmonic
functions.

The functions $\Phi_n(\rho,\Omega)$ are the
eigenfunctions of the angular part of the Faddeev equations, and the
radial coefficients $f_n(\rho)$ are obtained from the coupled set of
radial equations where the eigenvalues of the angular part enter as
effective adiabatic potentials \cite{nie01}.  Due to the complex
scaling the wave functions fall off exponentially at large hyperradii
for bound states and resonances with width-to-energy ratios
less than twice the rotation angle.

The components of the solution directly contain information about
the symmetry and hence about the isospin mixing, which can vary with
hyperradius.  This point has apparently not been appreciated in the
old analyses \cite{deb71,wit75}.  For instance, in $^6$Li
($n$+$p$+$\alpha$) the total isospin is
obtained by coupling neutron and proton isospins to either 0 or 1.  
Components with both isospins are in principle permitted,  
like for instance $(\ell_x,\ell_y,L,S,J)=(0,2,2,0,2),(1,1,2,0,2)$,
where $\ell_x$ and $\ell_y$ are orbital angular momenta related to 
the Jacobi coordinates $\bm{x}$ and $\bm{y}$ (with $\bm{x}$ proportional 
to the neutron-proton distance), and $L$, $S$ and $J$ are the total orbital, spin and
total angular momenta. In fact these two components do contribute, and are coupled 
due to the presence of the $\alpha$-particle, which Coulomb interacts with the proton 
but not with the neutron. This can totally break the (isospin) symmetry if the
symmetry breaking interaction is dominating as it is for large
distances. Thus, small isospin mixing at small distance can be
compatible with a very large isospin mixing at large distance.

The kinetic energy distribution of the fragments after decay of a
resonance is, except for a phase-space factor, obtained as the
absolute square of the total wave function in coordinate space for a
large value of the hyperradius $\rho$, but where the five hyperangles
are interpreted as in momentum space \cite{gar05c,gar05f}.
After integration over the four hyperangles $(\Omega_x,\Omega_y)$
describing the directions of the two Jacobi momenta, $\bm{k}_x$ and
$\bm{k_y}$, conjugate to $\bm{x}$ and $\bm{y}$, the probability
distribution as function of $k_y^2 \propto \cos^2 \alpha$, where
$\alpha$ is the fifth momentum hyperangle, is given by
\begin{equation}
 P(k_y^2) \propto P(\cos^2\alpha) \propto \sin(2\alpha)
    \int d\Omega_x d\Omega_y|\Psi(\rho,\alpha,\Omega_x,\Omega_y)|^2 \;.
\label{eq2}
\end{equation}
The kinetic energy of the third particle is proportional to $k_y^2
\propto \cos^2 \alpha$ which then gives the energy of the particle
relative to its maximum possible energy in the decay process.  These
observables carry information about initial state and decay mechanisms.

\paragraph*{Details of the calculations.}

The 2$^+$ resonance in $^6$Li has been computed using the same $\alpha$-nucleon 
interaction as in \cite{cob98} for $^6$He. For the
neutron-proton potential we use the one in \cite{cob97}. Components with relative
two-body orbital angular momenta up to 4 are considered. The main components 
included in the calculation are shown in table \ref{tab1} (the first column
labels the components). A proper choice of the
maximum value of the hypermomentum ($K_{max}$) for each of them is crucial
to obtain a correct convergence of the effective potentials at a sufficiently large 
distance. The $K_{max}$ value for the components 6 to 11 in the left part of table 
\ref{tab1} is relatively large to ensure an accurate calculation of their contribution, 
since these are precisely the components with zero isospin ($T$ in the table) that are not allowed in 
$^6$He or $^6$Be. For the remaining components (not shown in the table) the $K_{max}$-value 
is at least 20.

\begin{table}
\caption{Components included for the $2^+$-state in $^6$Li. 
The left part refers to the first Jacobi set ($\bm{x}$ from neutron to proton), 
and the right part to the second and third Jacobi sets
($\bm{x}$ from one of the nucleons to the $\alpha$-particle). The first
column numbers the different components.}
\label{tab1}
\vspace*{0.5cm}
\begin{center}
\begin{tabular}{|c|ccccccc|cccccc|}
\hline
  &\multicolumn{7}{|c|}{1$^{st}$ Jacobi set} & 
        \multicolumn{6}{c|}{2$^{nd}$ and 3$^{rd}$ Jacobi sets} \\
  &\multicolumn{7}{|c|}{
\put(33,3){\circle*{7}} \put(64,3){\circle*{7}}  \put(33,3){\vector(1,0){27}}
\put(21,3){$N$} \put(68,3){$N$} \put(43,5){$\vec{x}$}  
\put(48,-25){\circle*{13}} \put(58,-25){$\alpha$} 
\put(48,-25){\vector(0,1){27}}  \put(51,-12){$\vec{y}$}
                         } & 
  \multicolumn{6}{c|}{
\put(33,3){\circle*{13}} \put(65,3){\circle*{7}}  \put(33,3){\vector(1,0){27}}
\put(18,3){$\alpha$} \put(69,3){$N$} \put(46,5){$\vec{x}$} 
\put(50,-25){\vector(-1,4){6.5}}  \put(50,-12){$\vec{y}$}
\put(50,-25){\circle*{7}} \put(57,-25){$N$}
                       } \\  
  &  &  &  &  &  &  &  &  &  &  &  &  &  \\ \hline
   & $\ell_x$ & $\ell_y$ & $L$ & $s_x$ & $S$ & $K_{max}$ &  $T$  &
    $\ell_x$ & $\ell_y$ & $L$ & $s_x$ & $S$ & $K_{max}$ \\ \hline
1  & 0 & 2 & 2 & 0 & 0 & 240& 1 & 0 & 2 & 2 & 1/2 & 0 & 44  \\
2  & 2 & 0 & 2 & 0 & 0 & 180& 1 & 0 & 2 & 2 & 1/2 & 1 & 44  \\
3  & 1 & 1 & 1 & 1 & 1 & 180& 1 & 2 & 0 & 2 & 1/2 & 0 & 70  \\
4  & 1 & 1 & 2 & 1 & 1 & 64 & 1 & 2 & 0 & 2 & 1/2 & 1 & 44  \\
5  & 2 & 2 & 2 & 0 & 0 & 90 & 1 & 1 & 1 & 1 & 1/2 & 1 & 240 \\
6  & 0 & 2 & 2 & 1 & 1 & 240& 0 & 1 & 1 & 2 & 1/2 & 0 & 240 \\
7  & 2 & 0 & 2 & 1 & 1 & 240& 0 & 1 & 1 & 2 & 1/2 & 1 & 44  \\
8  & 1 & 1 & 2 & 0 & 0 & 240& 0 & 2 & 2 & 1 & 1/2 & 1 & 32  \\
9  & 2 & 2 & 1 & 1 & 1 & 240& 0 & 2 & 2 & 2 & 1/2 & 0 & 50  \\
10 & 2 & 2 & 2 & 1 & 1 & 240& 0 & 2 & 2 & 2 & 1/2 & 1 & 42  \\
11 & 2 & 2 & 3 & 1 & 1 & 240& 0 & 1 & 3 & 2 & 1/2 & 0 & 42  \\ \hline
\end{tabular} 
\end{center}
\end{table}

\begin{figure}
\vspace*{-1.cm}
\centerline{\psfig{figure=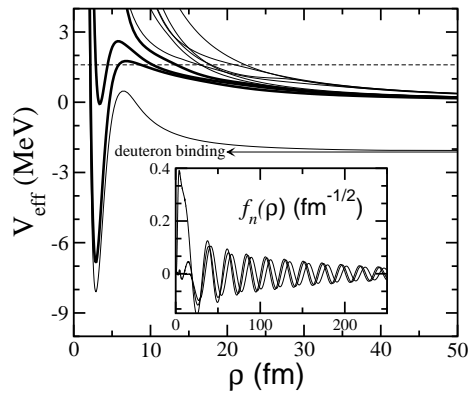,angle=270,width=10cm}}
\vspace*{-2.cm}
\caption{Outer panel: Real parts of the (complex scaled) effective adiabatic potentials for the 
2$^+$ resonance in $^6$Li. Inner panel: Real parts of the radial wave functions associated to the three most 
contributing effective potentials (indicated by the thick curves in the outer part).} 
\label{fig1}
\end{figure}

The outer part of Fig.\ref{fig1} shows the real parts of the effective potentials obtained after use of the 
(complex scaled) hyperspheric adiabatic expansion method. The lowest effective potential converges towards
the deuteron binding energy, and appears due to the inclusion of the components with zero
isospin in the neutron-proton channel. These potentials are indistinguishable
from the ones obtained when the basis size is reduced by a factor of two. This fact guaranties the convergence
of the potentials at least up to 100 fm. 

Calculation of the resonance wave function requires specification 
of the corresponding boundary condition. As shown in \cite{gar07}, a simple box boundary condition at a sufficiently 
large distance is enough to obtain a resonance wave function with the proper asymptotics. This requires 
extrapolation of the effective potentials for $\rho$ values beyond 100 fm. An expansion of the form 
$A/\rho+B/\rho^2+C/\rho^3+\cdots$ is
used (except for the lowest one). The non-adiabatic coupling functions $P_{nn^\prime}(\rho)$ and
$Q_{nn^\prime}(\rho)$ (see \cite{nie01}) are extrapolated as
$Q_{nn}(\rho)=A_{Q_{nn}}/\rho^2 + B_{Q_{nn}}/\rho^3 + C_{Q_{nn}}/\rho^4$ for diagonal $Q's$ (diagonal
$P's$ are zero), and as $A_{P,Q}/\rho^3 + B_{P,Q}/\rho^4 + C_{P,Q}/\rho^5$ for the non-diagonal $P$'s and $Q$'s.

A box boundary condition at $\rho_{max}$=1000 fm gives rise to a 2$^+$ 
resonance in $^6$Li with 
energy and width ($E_R$,$\Gamma_R$)=(1.67,0.51) MeV (5.37 MeV excitation energy), that agrees with 
the experimental value of ($E_R$,$\Gamma_R$)=(1.67$\pm$0.02,0.54$\pm$0.02) MeV  \cite{ajz88} 
(the resonance energy is indicated in Fig.\ref{fig1}  by the dashed line). However, the extrapolation 
used for the effective potentials implies that the asymptotics  
of the radial wave functions must be $F_{\xi}(\eta,\kappa\rho)-iG_{\xi}(\eta,\kappa\rho)$, where 
$\kappa=\sqrt{2mE/\hbar^2}$, $F_{\xi}$ and $G_{\xi}$ are the regular and irregular Coulomb functions, 
and the Coulomb charge $\eta$ and the index $\xi$ can be easily obtained from $A$, $B$, and
$A_{Q_{nn}}$. When this asymptotic condition is imposed, a much smaller value of $\rho_{max}$ is 
enough to obtain the resonance wave function (but still $\rho_{max}>$100 fm, and the expansions of the 
effective potentials, $P's$, and $Q's$ are required). 

\begin{figure}
\vspace*{-0.5cm}
\centerline{\psfig{figure=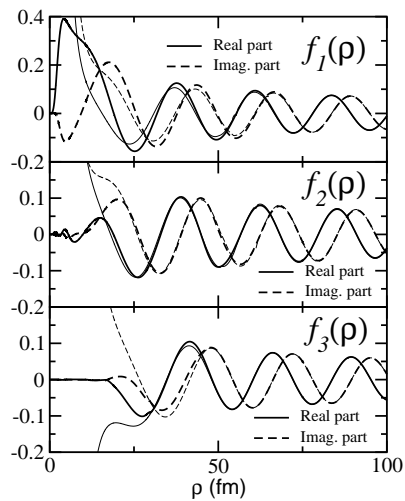,angle=270,width=10cm}}
\vspace*{-0.4cm}
\caption{Thick curves: Radial wave functions corresponding to the three most contributing effective
potentials. Thin curves: Expected asymptotics ($F_{\xi}(\eta,\kappa\rho)-iG_{\xi}(\eta,\kappa\rho)$) 
from the extrapolated potentials for $\rho>$100 fm.}
\label{fig2}
\end{figure}

 In the inner part of Fig.\ref{fig1} we show the real parts of the computed radial wave functions. A complex
scaling angle of 0.10 rads has been used. To make the picture cleaner we only show the ones associated with
the three most contributing effective potentials (thick curves in the outer part of the figure).
It is important to note that the contribution from the lowest adiabatic potential, the one holding deuteron 
at large distances, is very small, providing about $10^{-3}$\% of the norm, which is consistent with 
\cite{deb71,wit75}.
It is also necessary to investigate whether the radial wave functions have already reached the asymptotic
behaviour for $\rho < 100$ fm, i.e., in the $\rho$-region where the calculation is purely numerical. 
This would mean the asymptotics given by the extrapolated effective potentials, $P's$, and $Q's$,
is consistent with the numerical results.

This is tested in Fig.\ref{fig2}, where the real and imaginary parts of the
three most relevant radial wave functions (thick curves) are shown and compared to the expected asymptotics
as given by the regular and irregular Coulomb functions, with Coulomb charges and indices computed numerically
from the extrapolation of the potentials (thin curves). The matching between the
numerical wave functions and the asymptotics is already very good at about 60 fm, clearly below the
$\rho$-value (100 fm) from which the extrapolations are used. The labels 1, 2, and 3 refer to the deepest
thick potential, intermediate thick potential, and repulsive thick potential in the outer part of
Fig.\ref{fig1}.

  From the three-body wave function in Eq.(\ref{eq1}), we define the total weight (as function of $\rho$) as:
\begin{equation}
W(\rho)=\int \sin^2\alpha \cos^2\alpha d\alpha d\Omega_x d\Omega_y|\Psi(\rho,\alpha,\Omega_x,\Omega_y)|^2,
\label{eq3}
\end{equation}
When writing $\Psi$ in the first Jacobi set ($\bm{x}$ from neutron to proton) we find,
after integration over $\rho$ in Eq.(\ref{eq3}), that roughly 82\% of the weight is given by the three first 
components in the left part of table \ref{tab1} (38\%, 20\%, and 24\%), while the remaining 18\% is 
distributed among the other components. 

\begin{figure}
\vspace*{-0.3cm}
\centerline{\psfig{figure=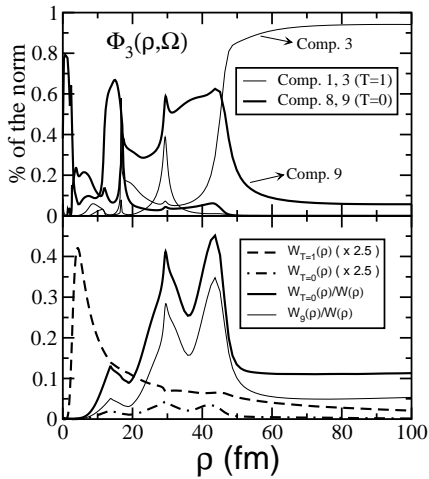,angle=270,width=10cm}}
\vspace*{-0.8cm}
\caption{Top: Main contributions (as a function of $\rho$) of the components in the left part of 
table \ref{tab1} to the eigenfunction $\Phi_3(\rho,\Omega)$ associated to the third adiabatic 
potential (thick repulsive potential in Fig.\ref{fig1}). The thin and thick curves correspond to components 
associated to $T$=1 and $T$=0, respectively. Bottom: Weight of the $T$=1 (dashed) 
and $T$=0 (dot-dashed) components in the 2$^+$ resonance wave function in $^6$Li (scaled by a factor 2.5). 
The thick and thin solid curves are the relative weight of the $T$=0 components and the relative weight of the 
9$^{th}$ component in the left part of table \ref{tab1}, respectively.}
\label{fig3}
\end{figure}

It is remarkable that the last components from 6 to 11 in the left part of table \ref{tab1} 
(associated to zero isospin) 
accumulate about 4.5\% of the integrated weight, most of it corresponding to component 9. 
This is due to the third adiabatic potential (repulsive thick potential in Fig.\ref{fig1}), whose 
corresponding eigenfunction $\Phi_3(\rho,\Omega)$ is dominated at intermediate distances by 
the components with zero isospin in the neutron-proton channel. This is shown in the upper part of 
Fig.\ref{fig3}, where we show, as a function of $\rho$, the main contributions to $\Phi_3(\rho,\Omega)$ 
from the components in the left part of table~\ref{tab1}. The thick curves correspond to components 
8 and 9 in the table ($T$=0). As observed in the figure, these components have a non-negligible 
weight at intermediate distances. In particular, component 9 gives a large contribution from 20 
to 50 fm. Beyond 50 fm component 3 dominates ($T$=1), but still a 5\% contribution from component 
9 is present. The rapid transition in $\Phi_3$ from component 9 to 3 reflects that the
isospin symmetry is totally broken, because only the Coulomb interaction is active. The lowest 
centrifugal barrier with $\ell_y$=1 is then abruptly preferred over $\ell_y$=2.
At short distances, as seen on the bottom of Fig.\ref{fig2}, the radial coefficient 
corresponding to the third adiabatic potential is negligible, but beyond 20 fm, the amplitude of 
$f_3(\rho)$ is similar to the one in $f_1(\rho)$ and $f_2(\rho)$. 

\begin{figure}[t]
\vspace*{-1.cm}
\centerline{\psfig{figure=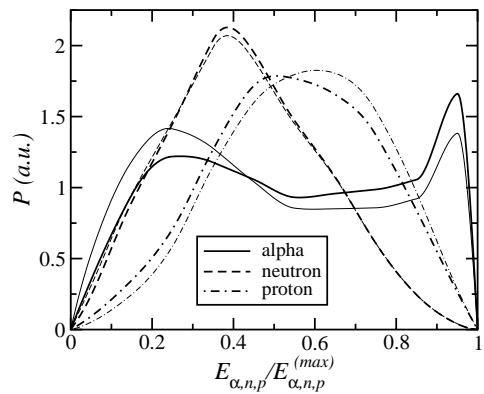,angle=270,width=10cm}}
\vspace*{-1.5cm}
\caption{ Thick curves: Energy distribution of the $\alpha$ (solid), neutron (dashed), and 
proton (dot-dashed) after decay of the 2$^+$ resonance in $^6$Li. Thin curves: The same 
energy distributions after excluding the $T$=0 components.}
\label{fig4}
\end{figure}

  Let us denote now by $W_{T=1}$ and $W_{T=0}$ the contributions to $W(\rho)$ in Eq.(\ref{eq3}) 
from the $T$=1 and $T$=0 components in the three-body wave function, respectively. These contributions
are shown (scaled by a factor 2.5) by the dashed and dot-dashed curves in the bottom part of Fig.\ref{fig3}. 
At short distances the $T$=1 contribution clearly dominates, while $W_{T=0}$ becomes relevant 
at intermediate $\rho$'s. This is more clearly seen by the thick solid curve, that shows the relative weight
of the $T$=0 components. From 20 to 50 fm this weight reaches up to 40\% of the total. This region 
coincides with the one where $\Phi_3(\rho,\Omega)$ has a relevant contribution from the $T$=0 components  
(upper part of the figure). Beyond 50 fm $W_{T=0}$ stabilizes at about 10\%.
In the figure the thin curve shows the relative contribution to the total weight from component 9.
This contribution gives most of the $T$=0 contribution, and governs the general
behaviour of $W_{T=0}/W$.

The non-negligible isospin mixing found at large distances can be relevant for those observables 
sensitive to the asymptotic behaviour of the wave function, like for instance the energy distributions
of the fragments after decay. In Fig.\ref{fig4} we show the $\alpha$ (solid), neutron (dashed),
and proton (dot-dashed) energy distributions according to Eq.(\ref{eq2}). The thick curves show the
total distributions, while the thin ones give the same distributions when
the $T$=0 components have been excluded. The results shown in the figure are very stable for $\rho$
values in Eq.(\ref{eq2}) ranging from 65 to 85 fm. In particular the curves shown in the figure
correspond to $\rho=75$ fm. As seen in the figure, inclusion of the $T$=0 components produce a visible
change in the energy distributions, although the general behaviour of the distributions does not change.

\paragraph*{Summary and conclusions.}

We have investigated dynamic isospin mixing in nuclear resonances. We
illustrate by the highest known  $2^+$ resonance of the three-body system $^6$Li,
where components with different isospin simultaneously can be present.
The Coulomb interaction between $\alpha$-particle and proton is
breaking the isospin symmetry and mixing isospins of 0 and 1.  
The isospin zero components essentially only appear in the neutron-proton 
continuum. The deuteron is populated in the decay by about $10^{-3}$\%,
that is consistent with the  experimental upper limits \cite{deb71,wit75}. 
The amount of isospin zero is consistent with that found in experiments 
\cite{wer81,gai88} (up to 30\% isospin mixing for low energies), while here 
the isospin mixing is due to the decay process and not direct reactions 
as suggested by these authors. 

We have used the complex scaled, hyperspheric adiabatic expansion
method with an extraordinary large basis for the components with zero
isospin.  The isospin content of the accurately computed resonance
wave functions vary substantially from small to large distances. The
relative isospin 0 contribution is small at small distances where the
main contribution resides. This relative contribution reaches about
40\% at intermediate distances, and stabilizes beyond 50 fm at roughly
10\% of the total. The total contribution integrated over all
distances of the isospin 0 components is about 4\% which is much
larger than for ordinary stable nuclei.  This mechanism of dynamic
isospin mixing is a common feature in decays of nuclear 
three- (or more-) body resonances.

\end{document}